\pdfoutput=1
\documentclass[12pt,a4paper]{article}
\usepackage{subcaption}
\usepackage{ifthen} 
\newboolean{pdflatex}
\setboolean{pdflatex}{true} 

\newboolean{articletitles}
\setboolean{articletitles}{true} 

\newboolean{uprightparticles}
\setboolean{uprightparticles}{false} 

\def\paperauthors{LHCb collaboration} 
\def\paperasciititle{Evidence for the decay Bs -> phi eta'} 
\def\papertitle{Evidence for the decay $\decay{\Bs}{\phi \eta'}$ } 
\def\paperkeywords{{High Energy Physics}, {LHCb}} 
\def\papercopyright{\the\year\ CERN for the benefit of the LHCb collaboration} 
\def\paperlicence{CC BY 4.0 licence}
\def\paperlicenceurl{https://creativecommons.org/licenses/by/4.0/}

\newif\ifEnableSectionTOCLinks
\EnableSectionTOCLinksfalse 

\usepackage[top=1in, bottom=1.25in, left=1in, right=1in]{geometry}

\usepackage{microtype}
\usepackage{lineno}  
\usepackage{xspace} 
\usepackage{caption} 

\usepackage{graphicx}  
\usepackage{color}
\usepackage{colortbl}
\graphicspath{{./figs/}} 

\usepackage{amsmath} 
\usepackage{amssymb}
\usepackage{amsfonts}
\usepackage{upgreek} 

\newcommand*\patchAmsMathEnvironmentForLineno[1]{%
\expandafter\let\csname old#1\expandafter\endcsname\csname #1\endcsname
\expandafter\let\csname oldend#1\expandafter\endcsname\csname
end#1\endcsname
 \renewenvironment{#1}%
   {\linenomath\csname old#1\endcsname}%
   {\csname oldend#1\endcsname\endlinenomath}%
}
\newcommand*\patchBothAmsMathEnvironmentsForLineno[1]{%
  \patchAmsMathEnvironmentForLineno{#1}%
  \patchAmsMathEnvironmentForLineno{#1*}%
}
\AtBeginDocument{%
\patchBothAmsMathEnvironmentsForLineno{equation}%
\patchBothAmsMathEnvironmentsForLineno{align}%
\patchBothAmsMathEnvironmentsForLineno{flalign}%
\patchBothAmsMathEnvironmentsForLineno{alignat}%
\patchBothAmsMathEnvironmentsForLineno{gather}%
\patchBothAmsMathEnvironmentsForLineno{multline}%
\patchBothAmsMathEnvironmentsForLineno{eqnarray}%
}

\usepackage[pdftex,
            pdfauthor={\paperauthors},
            pdftitle={\paperasciititle},
            pdfkeywords={\paperkeywords}]{hyperref}
\usepackage{hyperxmp}
\hypersetup{
    pdfcopyright={Copyright (C) \papercopyright},
    pdflicenseurl={\paperlicenceurl}
}

\usepackage[colorinlistoftodos,textsize=scriptsize]{todonotes}

\usepackage[bottom,flushmargin,hang,multiple]{footmisc}

\usepackage[all]{hypcap} 

\usepackage{xspace}
\usepackage{upgreek}

\def\lhcb   {\mbox{LHCb}\xspace}

\def\MagUp {\mbox{\em Mag\kern -0.05em Up}\xspace}

\ifthenelse{\boolean{uprightparticles}}%
{

 \def\Peta        {\ensuremath{\upeta}\xspace}

 \def\Pmu         {\ensuremath{\upmu}\xspace}

 \def\Ppi         {\ensuremath{\uppi}\xspace}
 
 \def\Prho        {\ensuremath{\uprho}\xspace}

 \def\Ppsi        {\ensuremath{\uppsi}\xspace}

 \def\PDelta      {\ensuremath{\Delta}\xspace}
 \def\PXi         {\ensuremath{\Xi}\xspace}
 \def\PLambda     {\ensuremath{\Lambda}\xspace}
 \def\PSigma      {\ensuremath{\Sigma}\xspace}
 \def\POmega      {\ensuremath{\Omega}\xspace}
 \def\PUpsilon    {\ensuremath{\Upsilon}\xspace}
 \let\oldPi\Pi
 \def\PPi         {\ensuremath{\oldPi}\xspace}

 \def\PB      {\ensuremath{\mathrm{B}}\xspace}
 \def\PD      {\ensuremath{\mathrm{D}}\xspace}
 \def\PJ      {\ensuremath{\mathrm{J}}\xspace}
 \def\PK      {\ensuremath{\mathrm{K}}\xspace}
 \def\Pb      {\ensuremath{\mathrm{b}}\xspace}
 \def\Pc      {\ensuremath{\mathrm{c}}\xspace}

 \def\Pp      {\ensuremath{\mathrm{p}}\xspace}

 \def\Ps      {\ensuremath{\mathrm{s}}\xspace}

 \def\thebaroffset{0.0em}
}
{

 \def\Peta        {\ensuremath{\eta}\xspace}

 \def\Pmu         {\ensuremath{\mu}\xspace}

 \def\Ppi         {\ensuremath{\pi}\xspace}
 
 \def\Prho        {\ensuremath{\rho}\xspace}

 \def\Ppsi        {\ensuremath{\psi}\xspace}
 
 \mathchardef\PDelta="7101
 \mathchardef\PXi="7104
 \mathchardef\PLambda="7103
 \mathchardef\PSigma="7106
 \mathchardef\POmega="710A
 \mathchardef\PUpsilon="7107
 \mathchardef\PPi="7105
 \def\PB      {\ensuremath{B}\xspace}
 \def\PD      {\ensuremath{D}\xspace}
 \def\PJ      {\ensuremath{J}\xspace}
 \def\PK      {\ensuremath{K}\xspace}
 \def\Pb      {\ensuremath{b}\xspace}
 \def\Pc      {\ensuremath{c}\xspace}

 \def\Pp      {\ensuremath{p}\xspace}

 \def\Ps      {\ensuremath{s}\xspace}

 \def\thebaroffset{0.18em}
}
\newcommand{\offsetoverline}[2][\thebaroffset]{\kern #1\overline{\kern -#1 #2}}%

\makeatletter
\ifcase \@ptsize \relax
  \newcommand{\miniscule}{\@setfontsize\miniscule{4}{5}}
\or
  \newcommand{\miniscule}{\@setfontsize\miniscule{5}{6}}
\or
  \newcommand{\miniscule}{\@setfontsize\miniscule{5}{6}}
\fi
\makeatother

\DeclareRobustCommand{\optbar}[1]{\shortstack{{\miniscule (\rule[.5ex]{1.25em}{.18mm})}
  \\ [-.7ex] $#1$}}

\def\mumu       {{\ensuremath{\Pmu^+\Pmu^-}}\xspace}

\def\squark    {{\ensuremath{\Ps}}\xspace}
\def\squarkbar {{\ensuremath{\overline \squark}}\xspace}

\def\cquark    {{\ensuremath{\Pc}}\xspace}

\def\bquark    {{\ensuremath{\Pb}}\xspace}
\def\bquarkbar {{\ensuremath{\overline \bquark}}\xspace}

\def\pion   {{\ensuremath{\Ppi}}\xspace}
\def\piz    {{\ensuremath{\pion^0}}\xspace}
\def\pip    {{\ensuremath{\pion^+}}\xspace}
\def\pim    {{\ensuremath{\pion^-}}\xspace}

\def\rhomeson {{\ensuremath{\Prho}}\xspace}
\def\rhoz     {{\ensuremath{\rhomeson^0}}\xspace}

\def\kaon    {{\ensuremath{\PK}}\xspace}

\def\KorKbar {\kern \thebaroffset\optbar{\kern -\thebaroffset \PK}{}\xspace}

\def\Kp      {{\ensuremath{\kaon^+}}\xspace}
\def\Km      {{\ensuremath{\kaon^-}}\xspace}

\def\Kstarz  {{\ensuremath{\kaon^{*0}}}\xspace}

\newcommand{\etapr}{\ensuremath{\Peta^{\prime}}\xspace}

\def\D       {{\ensuremath{\PD}}\xspace}

\def\DorDbar {\kern \thebaroffset\optbar{\kern -\thebaroffset \PD}\xspace}

\def\Dp      {{\ensuremath{\D^+}}\xspace}
\def\Dm      {{\ensuremath{\D^-}}\xspace}

\def\DpDm    {\ensuremath{\Dp {\kern -0.16em \Dm}}\xspace}

\def\Ds      {{\ensuremath{\D^+_\squark}}\xspace}

\def\B       {{\ensuremath{\PB}}\xspace}

\def\BorBbar {\kern \thebaroffset\optbar{\kern -\thebaroffset \PB}\xspace}

\def\Bd      {{\ensuremath{\B^0}}\xspace}

\def\BdorBdbar {\kern \thebaroffset\optbar{\kern -\thebaroffset \Bd}\xspace}
\def\Bu      {{\ensuremath{\B^+}}\xspace}

\def\Bs      {{\ensuremath{\B^0_\squark}}\xspace}

\def\BsorBsbar {\kern \thebaroffset\optbar{\kern -\thebaroffset \Bs}\xspace}

\def\Bds     {{\ensuremath{\B_{(\squark)}^0}}\xspace}

\def\BdorBs  {\Bds}

\def\jpsi     {{\ensuremath{{\PJ\mskip -3mu/\mskip -2mu\Ppsi}}}\xspace}

\def\Y#1S{\ensuremath{\PUpsilon{(#1S)}}\xspace}

\def\proton      {{\ensuremath{\Pp}}\xspace}

\def\Lz          {{\ensuremath{\PLambda}}\xspace}

\def\LorLbar     {\kern \thebaroffset\optbar{\kern -\thebaroffset \PLambda}\xspace}

\def\Lb           {{\ensuremath{\Lz^0_\bquark}}\xspace}

\newcommand{\decay}[2]{\mbox{\ensuremath{#1\!\to #2}}\xspace}

\def\to                 {\ensuremath{\rightarrow}\xspace}

\def\CP                {{\ensuremath{C\!P}}\xspace}

\def\AT#1     {\ensuremath{A_{\mathrm{T}}^{#1}}\xspace}           

\def\C#1      {\ensuremath{\mathcal{C}_{#1}}\xspace}                       
\def\Cp#1     {\ensuremath{\mathcal{C}_{#1}^{'}}\xspace}                    
\def\Ceff#1   {\ensuremath{\mathcal{C}_{#1}^{\mathrm{(eff)}}}\xspace}        
\def\Cpeff#1  {\ensuremath{\mathcal{C}_{#1}^{'\mathrm{(eff)}}}\xspace}       
\def\Ope#1    {\ensuremath{\mathcal{O}_{#1}}\xspace}                       
\def\Opep#1   {\ensuremath{\mathcal{O}_{#1}^{'}}\xspace}                    

\newcommand{\aunit}[1]{\ensuremath{\text{\,#1}}}

\newcommand{\tev}{\aunit{Te\kern -0.1em V}\xspace}
\newcommand{\gev}{\aunit{Ge\kern -0.1em V}\xspace}
\newcommand{\mev}{\aunit{Me\kern -0.1em V}\xspace}
\newcommand{\kev}{\aunit{ke\kern -0.1em V}\xspace}
\newcommand{\ev}{\aunit{e\kern -0.1em V}\xspace}

\newcommand{\mevc}{\ensuremath{\aunit{Me\kern -0.1em V\!/}c}\xspace}
\newcommand{\gevc}{\ensuremath{\aunit{Ge\kern -0.1em V\!/}c}\xspace}
\newcommand{\mevcc}{\ensuremath{\aunit{Me\kern -0.1em V\!/}c^2}\xspace}
\newcommand{\gevcc}{\ensuremath{\aunit{Ge\kern -0.1em V\!/}c^2}\xspace}

\def\fb   {\ensuremath{\aunit{fb}}\xspace}
\def\invfb   {\ensuremath{\fb^{-1}}\xspace}

\def\ps   {\ensuremath{\aunit{ps}}\xspace}

\def\gsim{{~\raise.15em\hbox{$>$}\kern-.85em
          \lower.35em\hbox{$\sim$}~}\xspace}
\def\lsim{{~\raise.15em\hbox{$<$}\kern-.85em
          \lower.35em\hbox{$\sim$}~}\xspace}

\def\pt         {\ensuremath{p_{\mathrm{T}}}\xspace}

\def\ptot       {\ensuremath{p}\xspace}

\def\evtgen     {\mbox{\textsc{EvtGen}}\xspace}

\def\geant      {\mbox{\textsc{Geant4}}\xspace}

\def\photos     {\mbox{\textsc{Photos}}\xspace}

\def\pythia     {\mbox{\textsc{Pythia}}\xspace}

\def\tell1  {TELL1\xspace}
\def\ukl1   {UKL1\xspace}

\newcommand{\lhcborcid}[1]{\href{https://orcid.org/#1}{\hspace*{0.1em}\raisebox{-0.45ex}{\includegraphics[width=1em]{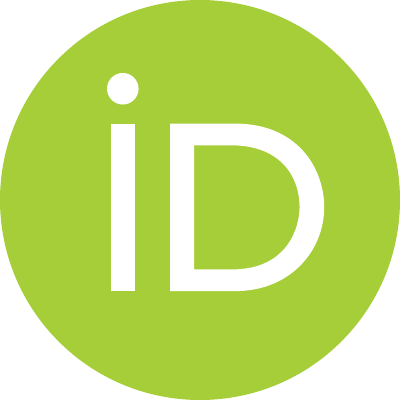}}}}

\hypersetup{
  colorlinks   = true, 
  urlcolor     = blue, 
  linkcolor    = blue, 
  citecolor    = red   
}

\ifEnableSectionTOCLinks
    \usepackage[explicit]{titlesec} 
    
    \let\oldcontentsline\contentsline
    \renewcommand

    \titleformat{\section}{\normalfont\Large\bf}{\hyperlink{tocsection.\thesection}{{\thesection} \parbox[t]{\dimexpr\textwidth-1pc}{#1}}}{1pc}{}

    \titleformat{\subsection}{\normalfont\bf}{\hyperlink{tocsubsection.\thesubsection}{{\thesubsection} \parbox[t]{\dimexpr\textwidth-1pc}{#1}}}{1pc}{}

    \titleformat{name=\section,numberless}[display]{}{}{0pt}{\normalfont\Huge\bfseries #1}
\fi

\usepackage{cite} 
\usepackage{mciteplus}

\usepackage{longtable} 

\begin{document}

\renewcommand{\thefootnote}{\fnsymbol{footnote}}
\setcounter{footnote}{1}


\begin{titlepage}
\pagenumbering{roman}

\vspace*{-1.5cm}
\centerline{\large EUROPEAN ORGANIZATION FOR NUCLEAR RESEARCH (CERN)}
\vspace*{1.5cm}
\noindent
\begin{tabular*}{\linewidth}{lc@{\extracolsep{\fill}}r@{\extracolsep{0pt}}}
\ifthenelse{\boolean{pdflatex}}
{\vspace*{-1.5cm}\mbox{\!\!\!\includegraphics[width=.14\textwidth]{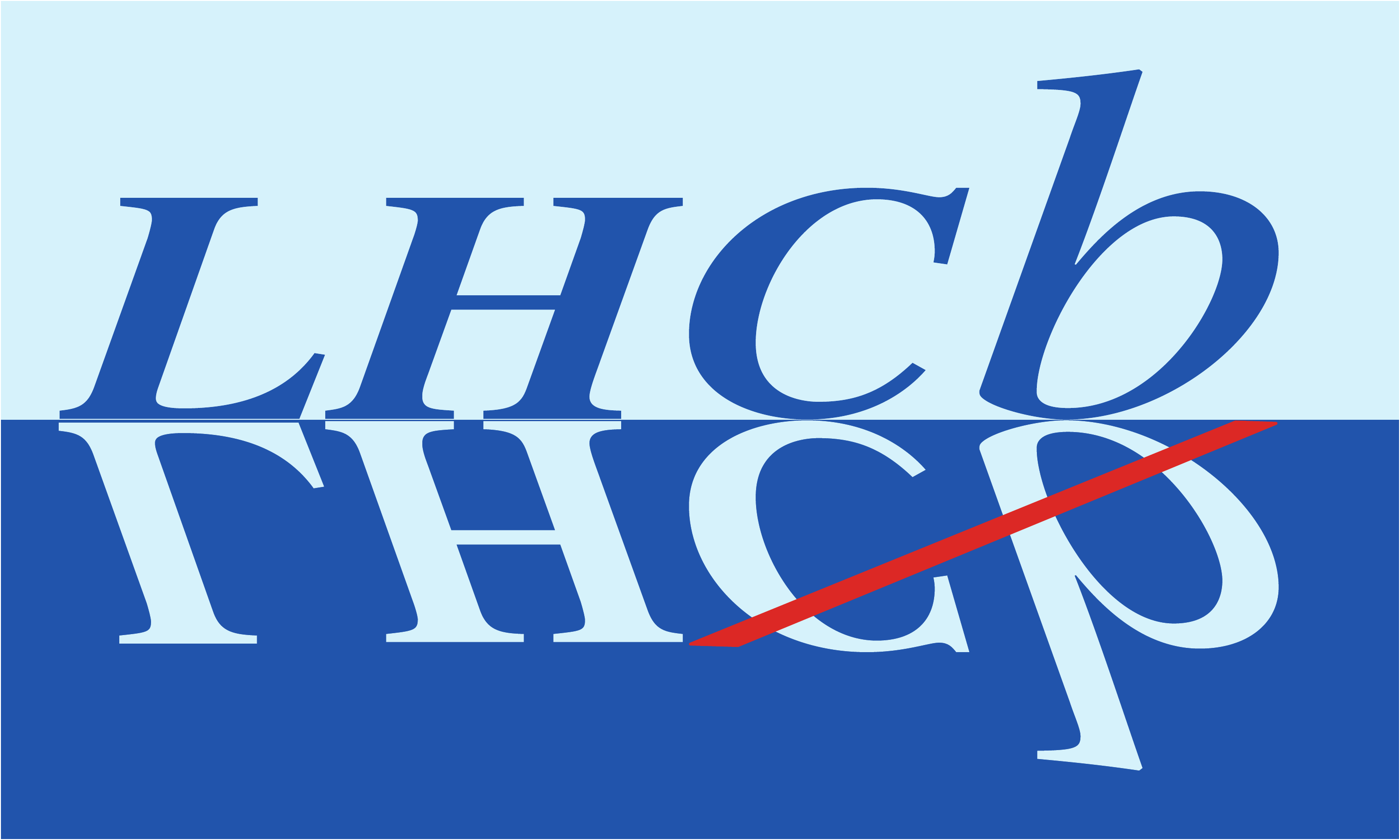}} & &}%
{\vspace*{-1.2cm}\mbox{\!\!\!\includegraphics[width=.12\textwidth]{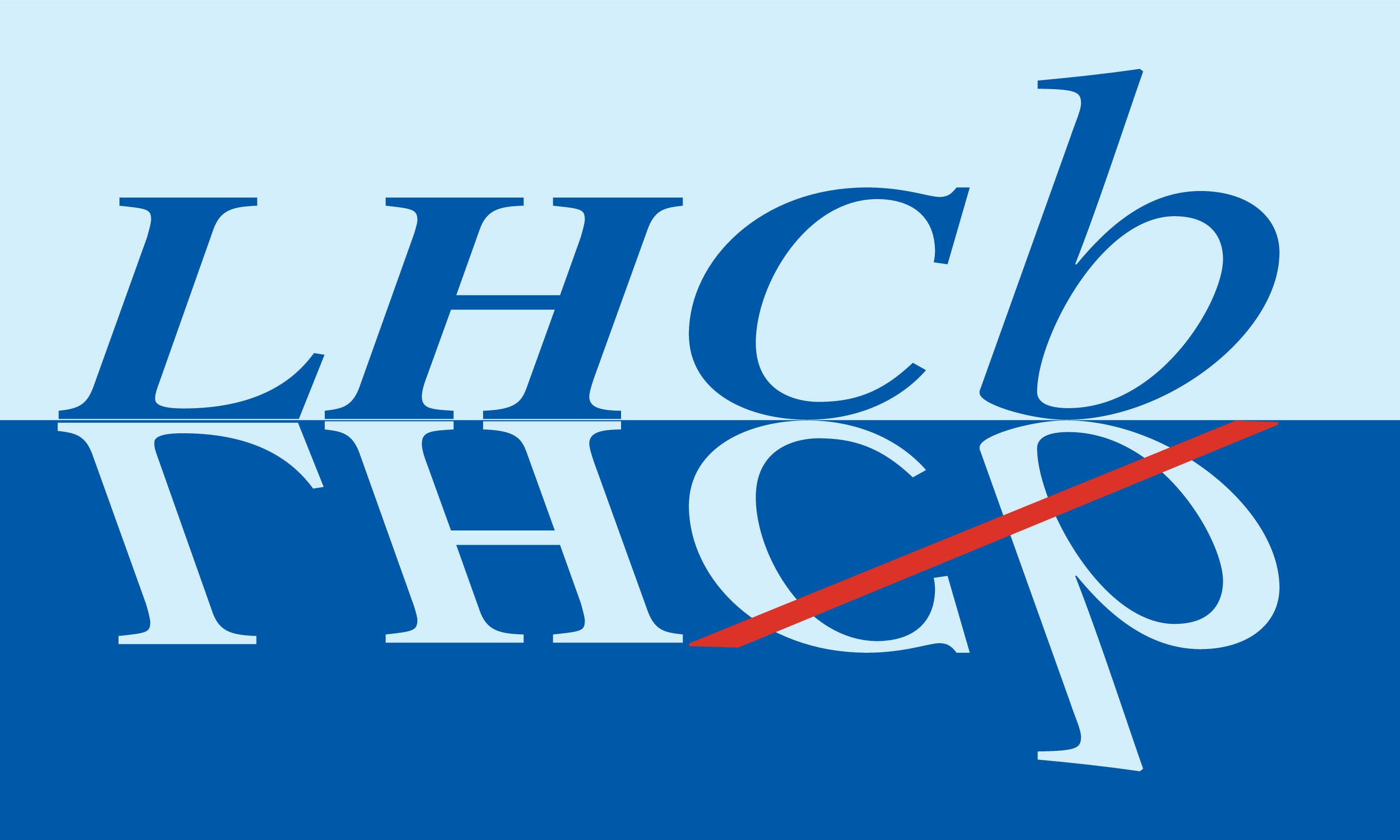}} & &}%
\\
 & & CERN-EP-2026-104 \\  
 & & LHCb-PAPER-2026-002 \\  
 & & May 8, 2026 \\ 
 & & \\
\end{tabular*}

\vspace*{4.0cm}

{\normalfont\bfseries\boldmath\huge
\begin{center}
  \papertitle 
\end{center}
}

\vspace*{2.0cm}

\begin{center}
\paperauthors\footnote{Authors are listed at the end of this paper.}
\end{center}

\vspace{\fill}

\begin{abstract}
  \noindent
 Using a dataset corresponding to an integrated luminosity of $9 \,\textrm{fb}^{-1}$ collected in proton-proton collisions between 2011 and 2018 by the LHCb experiment, evidence is found for the decay $\decay{\Bs}{\phi \eta'}$ with $3.5 \sigma$ significance. The branching ratio relative to the $\decay{\Bs}{\phi \phi}$ decay is determined to be $\mathcal{R}=(3.56 \pm 0.79\pm 0.18\pm 0.06)\times10^{-2}$. This corresponds to a branching fraction,  
 $\mathcal{B}(\decay{\Bs}{\phi \eta'})=(0.66 \pm 0.15 \pm 0.03 \pm 0.02) \times 10^{-6}$ where, in both cases, the first uncertainty is statistical, the second systematic, and the third due to external branching fractions.

\end{abstract}

\vspace*{2.0cm}

\begin{center}
  Published in JHEP 10(2026)055
 
\end{center}

\vspace{\fill}

{\footnotesize 
\centerline{\copyright~\papercopyright. \href{\paperlicenceurl}{\paperlicence}.}}
\vspace*{2mm}

\end{titlepage}


\newpage
\setcounter{page}{2}
\mbox{~}
%
%
%
%


\renewcommand{\thefootnote}{\arabic{footnote}}
\setcounter{footnote}{0}


\cleardoublepage


\pagestyle{plain} 
\setcounter{page}{1}
\pagenumbering{arabic}


\section{Introduction}
\label{sec:Introduction}
In the Standard Model (SM), $b$-hadron decays to charmless final states are highly suppressed due to the structure of the CKM matrix \cite{Kobayashi:1973fv}. Measurements of their branching fractions provide important tests for QCD-based descriptions of hadronic \B-meson decays. Theoretical predictions for charmless decay modes are made within the framework of perturbative QCD \cite{Ali:2007ff}, soft collinear effective theory (SCET) \cite{Wang:2008rk}, and QCD factorisation \cite{Cheng:2009mu}. These predictions have large uncertainties due to the complexity of calculating contributions from electroweak loop (penguin) diagrams. Analyses of charge-parity (\CP) asymmetries and ratios in these high-yield modes can provide more sensitive probes of contributions from physics beyond the SM.

\begin{figure}[b!]
    \centering
    \includegraphics[width=0.65\linewidth]{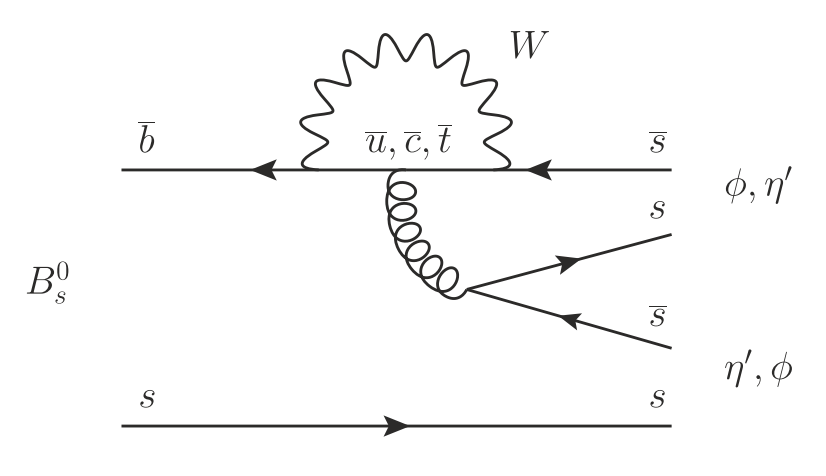}
    \caption{Feynman diagram for the $\decay{\Bs}{\phi \eta'}$ decay.  }
    \label{fig:feyn}
\end{figure}

The decay $\decay{\Bs}{\phi \eta'}$ proceeds through a $\decay{\bquarkbar}{\squarkbar \squark \squarkbar}$ transition (Fig.~\ref{fig:feyn}).\footnote{The symbols $\phi$ and $\etapr$ indicate the $\phi(1020)$ and $\etapr(958)$ mesons throughout.} Predictions for the branching fraction for this mode lie in the range $(0.05\--20)\times10^{-6}$ \cite{Beneke:2003zv,Cheng:2009mu,Ali:2007ff,Chen:2007qm, Wang:2008rk, PhysRevD.91.014011, Zhou:2016jkv}. This large range reflects the limited knowledge of form factors, penguin contributions, and $\omega\--\phi$ mixing.  An additional source of uncertainty arises from the fact that the spectator $\squark$~quark can hadronise to either the vector (V) meson $\phi$ or the pseudoscalar (P) meson $\eta'$ final-state, as shown in Fig.~\ref{fig:feyn}. Depending on the $\decay{\Bs}{\phi}$ form factor, there can be strong cancellations between the different P-V and V-P amplitudes, depending on which meson the spectator $\squark$~quark goes into, contributing to the $\decay{\Bs}{\phi \eta'}$ final state. This is not present in V-V modes such as $\decay{\Bs}{\phi \phi}$. Knowledge of the $\decay{\Bs}{\phi \eta'}$ branching fraction, as well as that of the related decay $\decay{\Bs}{\phi \eta}$, will discriminate between the QCD models and quantify the influence of the colour-suppressed QCD penguin loop \cite{Chen:2007qm}.

Experimentally, this decay was first searched for by the LHCb collaboration using $pp$ collisions at $\sqrt{s} = 7$ and $8\tev$ collected during the Run~1 of the LHC (2011--2012) and corresponding to an integrated luminosity of $3\invfb$~\cite{LHCB-PAPER-2016-060}. No signal was observed, and the limit of $\mathcal{B}(\decay{\Bs}{\phi \eta'}) < 0.82 \, (1.01) \times 10^{-6}$ was set at the $90\% \,(95\%) $ confidence level. This paper presents an updated study that includes data collected during Run~2 (2015–2018) at $\sqrt{s}=13 \tev$, yielding a total integrated luminosity of $9\invfb$, with improved detector calibration, alignment, and event selection. The result presented here supersedes the previous LHCb search and upper limit reported in Ref.~\cite{LHCB-PAPER-2016-060}. The first evidence for the decay $\decay{\Bs}{\phi \eta'}$ with $\decay{\phi}{\Kp \Km}$ and $\decay{\eta'}{\rhoz \gamma}$ is found. Its branching fraction is determined relative to the normalisation channel \decay{\Bs}{\phi\phi}, with \decay{\phi}{\Kp\Km}.

\section{Detector and simulation}
\label{sec:Detector}
The \lhcb detector~\cite{LHCb-DP-2008-001,LHCb-DP-2014-002} is a
single-arm forward spectrometer that covers the~\mbox{pseudorapidity}
range $2<\eta <5$ and was designed for the study of particles containing
\bquark or \cquark quarks. The detector configuration used during Run 1 and Run 2~includes a high-precision tracking
system consisting  of a silicon-strip vertex detector
surrounding the proton-proton ($\proton\proton$)~interaction region, a large-area
silicon-strip detector located upstream of a dipole magnet with 
bending power of approximately $4{\mathrm{\,T\,m}}$ \cite{LHCb-TDR-001}, and three stations of 
silicon-strip detectors and straw drift tubes  placed downstream of the~magnet. 
The~tracking system provides a measurement of the momentum, \ptot,  
of charged particles with a relative uncertainty that varies from~0.5\% at low momentum  
to~1.0\% at $p=200\gevc$.  Large samples of  $\decay{\jpsi}{\mumu}$~and
$\decay{\Bu}{\jpsi\Kp}$ decays,\footnote{The inclusion of charge-conjugate processes is implied throughout.} collected concurrently with
the dataset used in this analysis are used to calibrate the momentum
scale of the spectrometer~\cite{LHCb-DP-2023-003}. The relative uncertainty achieved on the momentum scale is $3 \times 10^{-4}$. 

Different types of charged hadrons are distinguished using information from two ring\nobreakdash-imaging 
Cherenkov\,(RICH) detectors. In addition, photons, electrons, and hadrons are 
identified by a calorimeter system consisting of scintillating-pad and preshower 
detectors, an electromagnetic calorimeter, and a hadronic calorimeter. Muons are
identified by a system composed of alternating  layers of iron and multiwire proportional chambers.

The online event selection is performed by a~trigger,  which consists
of a~hardware stage followed by a~two\nobreakdash-level software
stage~\cite{LHCb-DP-2012-004,LHCb-DP-2019-001}. At the hardware
trigger stage, events are required to contain a muon with high transverse momentum, \pt, 
or a hadron, photon, or electron with high transverse energy in the
calorimeters. For hadrons,  the transverse energy threshold is
around 3.5\gev. The software trigger requires a two-, three-, or four-track
secondary vertex with a significant displacement from any primary $pp$
interaction vertex (PV) \cite{LHCb-DP-2014-001}. At least one charged particle must have $\pt > 1.6\gevc$ and be inconsistent with
originating from a primary vertex. A multivariate
algorithm~\cite{BBDT,LHCb-PROC-2015-018} is used for the identification of secondary
vertices consistent with the decay of a $b$ hadron. In the offline
selection, trigger signals are associated with reconstructed
particles. Selection requirements can therefore be made on the trigger
selection itself and on whether the decision was due to the signal candidate only (TOS) or other
particles produced in the $pp$ collision but independent of the signal itself (TIS). In this analysis, the dataset is split into four categories based on the run period and the hardware trigger type, reflecting differences in selection efficiency across these categories. Triggered data further undergoes a centralised offline processing step to deliver physics-analysis-ready data across the entire \lhcb physics programme~\cite{Stripping}.

Simulation is used to determine selection requirements and efficiencies, and to describe the
invariant-mass distribution of the signal candidates. In the simulation, $pp$ collisions are generated using
\pythia~\cite{Sjostrand:2007gs,*Sjostrand:2006za}
with a specific \lhcb configuration~\cite{LHCb-PROC-2010-056}. The decays
of unstable particles
are described by \evtgen~\cite{Lange:2001uf}, in which final-state
radiation is generated using \photos~\cite{davidson2015photos}.
The interaction of the generated particles with the detector and its
response are implemented using the \geant
toolkit~\cite{Allison:2006ve, *Agostinelli:2002hh} as described in Ref.~\cite{LHCb-PROC-2011-006}. The full detector simulation reproduces the mass resolution observed in the data for $\bquark$ hadron decays to an accuracy of $5\%$ or better. Moreover, several sources of background contributions are studied using the~{\sc{RapidSim}} fast-simulation package~\cite{Cowan:2016tnm}. 

\section{Selection}
\label{sec:selection}
The offline selection consists of a loose preselection, a further selection to suppress specific backgrounds, and a multivariate algorithm that efficiently suppresses the residual combinatorial background contribution while maintaining a high signal efficiency.

The selection starts from high-quality charged particles with $\pt > 600 \mevc$, which are displaced from any PV. Candidate $\phi$ mesons are formed by pairing up oppositely charged particles that are identified as kaons by the RICH detectors. The vertex quality of the $\phi$ candidate is required to be good, and the invariant mass must be within $15 \mevcc$ of the known $\phi$ mass \cite{PDG2024}. This latter requirement reduces the background contribution from $\decay{\Kstarz(892)}{\Kp \pim}$ decays, where the pion is misidentified as a kaon, by two orders of magnitude. Possible background contribution from the rare decay $\decay{\Lb}{p \Km \phi}$ with $\decay{\phi}{\pip\pim\piz}$ is suppressed by requiring each kaon to be inconsistent with the proton hypothesis. 

Candidates for the $\decay{\eta'}{(\decay{\rhoz(770)}{\pip \pim}) \gamma}$ decay are formed by associating a pair of oppositely charged particles, identified as pions, with a combined invariant mass within the $620$ to $920 \mevcc$ range of $\rhoz$ mass window, along with a neutral cluster in the electromagnetic calorimeter identified as a photon with $\pt > 400 \mevc$. The mass of the $\eta'$ candidate is required to be in the range $920$ to $1000 \mevcc$. 

The $\phi$ and $\eta'$ candidates are combined to create $\Bs$ candidates. Combinatorial background is suppressed by requiring that the $\Bs$ candidates have $\pt > 2.5 \gevc$ and is displaced from any PV. A constrained vertex fit \cite{Hulsbergen:2005pu} is applied to check that the four tracks originate from a common point and that the $\Bs$ candidate originates from the associated PV, chosen as the PV that best matches the flight direction of the $\Bs$ candidate. In this fit, the invariant mass of the $\pip \pim \gamma$ candidate is constrained to the known $\eta'$ mass, while that of the $\Kp\Km$ candidate is fixed to the known $\phi$ mass \cite{PDG2024}. The fit is required to converge, and a loose requirement is made on the vertex fit quality.

Several selection requirements are applied to suppress specific background contributions from other $\bquark$ hadron decays. The contribution from  combining a $\decay{\BdorBs}{\phi \pip \pim}$ decay with a random photon produced in the $pp$ interaction is suppressed by requiring that the invariant mass of the four particles $m(\Kp\Km\pip\pim)$ is less than $5250 \mevcc$. To suppress the decay $\decay{\Bd}{\phi \Kstarz}$, the kaon mass hypothesis is considered for the pion that is the most poorly identified by the RICH detectors. If 
the calculated $m(\Kp\Km\Kp\pim)$ mass is within $50 \mevcc$ of the known $\Bd$ mass, the candidate is only kept if tighter particle-identification (PID) criteria are met. The contribution from the $\decay{\Bs}{\phi \phi}$ decay, where one of the $\phi$ mesons decays to $\pip\pim\piz$, is suppressed by requiring $|m(\pip\gamma)- m(\pim\gamma)| < 250 \mevcc$. The contribution of partially reconstructed $\bquark$ hadron decays to $\Ds X$, with $\decay{\Ds}{\phi \pip}$ final state is suppressed by rejecting candidates where either of the $m(\Kp\Km\pip)$ or $m(\Kp\Km\pim)$ masses is within $25 \mevcc$ of the known $\Ds$ mass.

Finally, a multivariate analysis is performed to suppress the remaining combinatorial background contributions. A gradient-boosted decision tree classifier with $k$-folding is trained using the {\sc{XGBoost}} package \cite{Chen:2016:XST:2939672.2939785}. The training is performed on the combined dataset of the Run~1 and Run~2 data-taking periods, using 18 variables related to the candidate kinematics, topology of the tracks including the helicity, vertex quality, cone isolation of the $\bquark$ hadron candidate \cite{LHCb-PAPER-2016-054}, and photon identification. Simulated $\decay{\Bs}{\phi \eta'}$ decays are used as signal proxy, and the \Bs mass upper sideband in data serves as the combinatorial background proxy. The selection requirement based on the classifier response is optimised for each category; the figure-of-merit described in Ref.~\cite{Punzi:2003bu} is maximised with the parameter $a$ set to $5$, corresponding to a significance of $5\sigma$ for the observation of a signal.

For the normalisation mode, candidates for the $\decay{\Bs}{\phi \phi}$ decays are created by combining pairs of $\phi$ mesons with the same selection and vertex fitting adopted for the signal mode.

\section{Invariant-mass fit}
\label{sec:massfit}

The modelling of the invariant-mass distribution in signal and background is performed using the simulation datasets for Run~1 and Run~2. The baseline model used to describe the  $\decay{\Bs}{\phi \eta'}$ decay mode is a double-sided Crystal Ball (DSCB) function. This is a generalisation of the Crystal
Ball function \cite{Skwarnicki:1986xj} with power law tails on both sides of the peak. In the fit to the data, the tail parameters are fixed to the values found in the simulation, while the mean and resolution parameters are free to vary.

The remaining background contributions after the selection need to be modelled in the fit. The most important background source is the decay $\decay{\Bs}{\phi \phi}$, with the subsequent decays $\decay{\phi}{\Kp \Km}$ and $\decay{\phi}{\pip \pim (\decay{\piz}{\gamma \gamma})}$, which include an undetected photon. This gives a relatively broad distribution, peaking around $5300 \mevcc$, which is modelled by an asymmetric Gaussian function. The remaining contribution from the rare decay $\decay{\Lb}{p \Km \eta'}$ \cite{Timothy2018} with the proton misidentified as a kaon is modelled by a wide DSCB function that peaks around $5500 \mevcc$. Finally, an exponential function is used to describe the combinatorial background. 

For the fit to the normalisation mode, a skewed Student's t-distribution \cite{jbes.2009.06161,mnsc.44.12.1650} is used to describe the $\decay{\Bs}{\phi \phi}$  signal. Combinatorial background is modelled using an exponential function. An asymmetric Gaussian function is used to describe the background from the decay $\decay{\Lb}{p \Km \phi}$ with a misidentified proton. This gives a broad distribution peaking around $5460 \mevcc$ \cite{LHCb-PAPER-2025-018}.

\section{Results}
\label{sec:results}
\begin{figure}[tb!]
    \centering
    \includegraphics[width=0.8\linewidth]{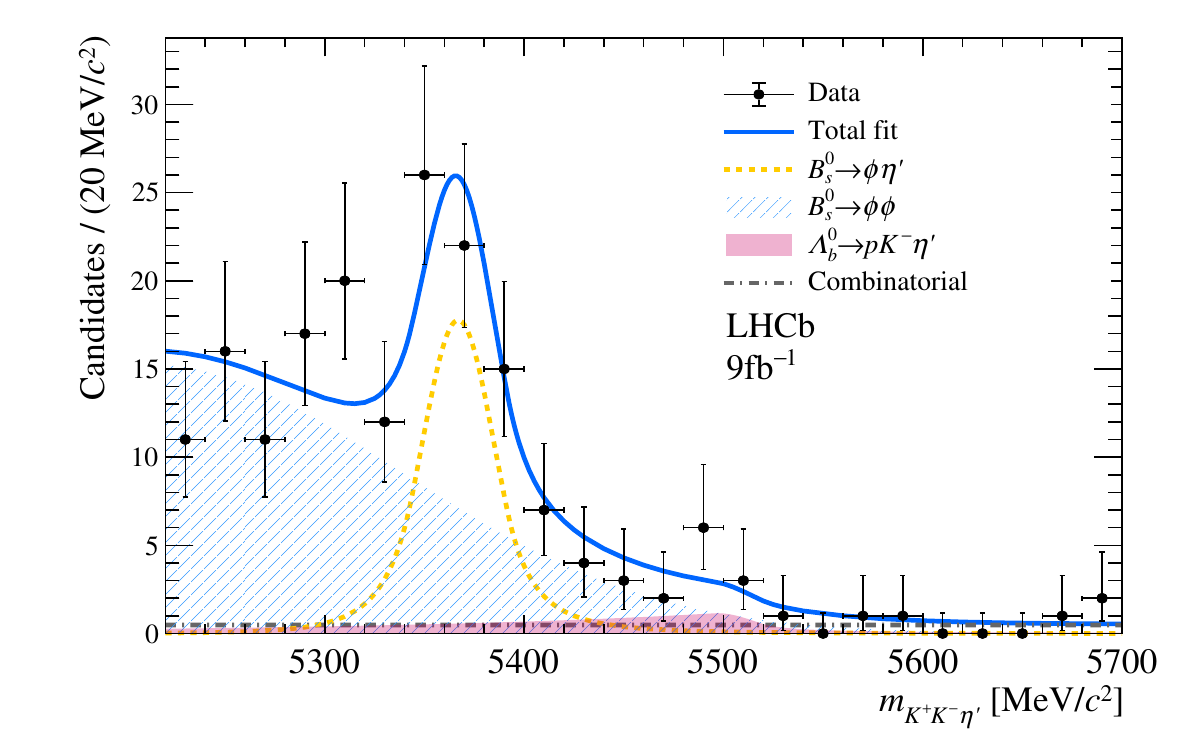}
    \caption{Distribution of the $K^+K^-\eta'$ invariant-mass for the complete dataset, summing over the four categories. The result of the simultaneous fit described in the text is also shown, with the signal component represented in yellow.}
    \label{fig:signalmass_total}
\end{figure}

To determine the $\decay{\Bs}{\phi \eta'}$ signal yield, a simultaneous unbinned maximum-likelihood fit is performed on the mass distribution of the selected signal candidates using the model described in Sect.~\ref{sec:massfit}. Figure~\ref{fig:signalmass_total} shows the invariant-mass distribution for the selected $K^+K^-\eta'$ candidates after all the selection requirements have been applied. 

In the signal fit, the mean and width of the DSCB function are shared among the four categories. For each category, the expected yield of the partially reconstructed $\decay{\Bs}{\phi \phi}$ background is determined from the observed yield of the normalisation mode and the relative efficiency obtained from simulation. This estimate is incorporated into the fit as a Gaussian constraint. The background from the decay $\decay{\Lb}{p \Km \eta'}$ is omitted from the Run~1 TIS category, which contains the smallest number of candidates. The signal yield is allowed to fluctuate to negative values. 

Pseudoexperiments are used to validate the fit procedure. In these studies, while the reported central values of the yields are reasonable, the associated uncertainties are underestimated, in particular for the Run~1 TIS category, which has the lowest data yield. Therefore, the uncertainty estimated from pseudoexperiments is used when calculating the final results. The fit yields and uncertainties obtained in this way are summarised in Table~\ref{tab:phieta_yield}. Summing across the four categories, the total signal yield is $46.4\pm9.7$. The significance of the signal, computed using Wilks' theorem \cite{Wilks:1938dza}, is $3.5 \sigma$ considering the statistical uncertainties and remains unchanged with systematics included. 

\begin{table}[tb!]
    \centering
     \caption{Fitted yields of the signal mode, together with the yields of the background components in Run~1 and Run~2. The uncertainties for the signal mode are estimated using pseudoexperiments to ensure proper coverage.}
    \begin{tabular}{l r@{\:$\pm$\:}l r@{\:$\pm$\:}l r@{\:$\pm$\:}l}
    \hline
               &\multicolumn{2}{c}{$N_{\phi\eta'}$} &\multicolumn{2}{c}{$N_{\phi\phi}$} &\multicolumn{2}{c}{$N_{\Lb}$} \\
    \hline
         Run~1 TIS & $-0.1$ & $2.6$ & $9.1$ & $3.6$ & \multicolumn{2}{c}{\textemdash}\\
         Run~1 TOS & $5.2$ & $3.1$ & $11.8$ & $1.8$ & $2.4$ & $0.4$  \\
         Run~2 TIS & $18.1$ & $6.0$ & $53.3$ & $3.7$ & $3.8$ & $0.9$  \\
         Run~2 TOS & $23.2$ & $6.5$ & $41.8$ & $3.7$ & $2.8$ & $1.0$  \\
    \hline
    \end{tabular}
    \label{tab:phieta_yield}
\end{table}

\begin{figure}[tb!]
    \centering
    \includegraphics[width=0.8\linewidth]{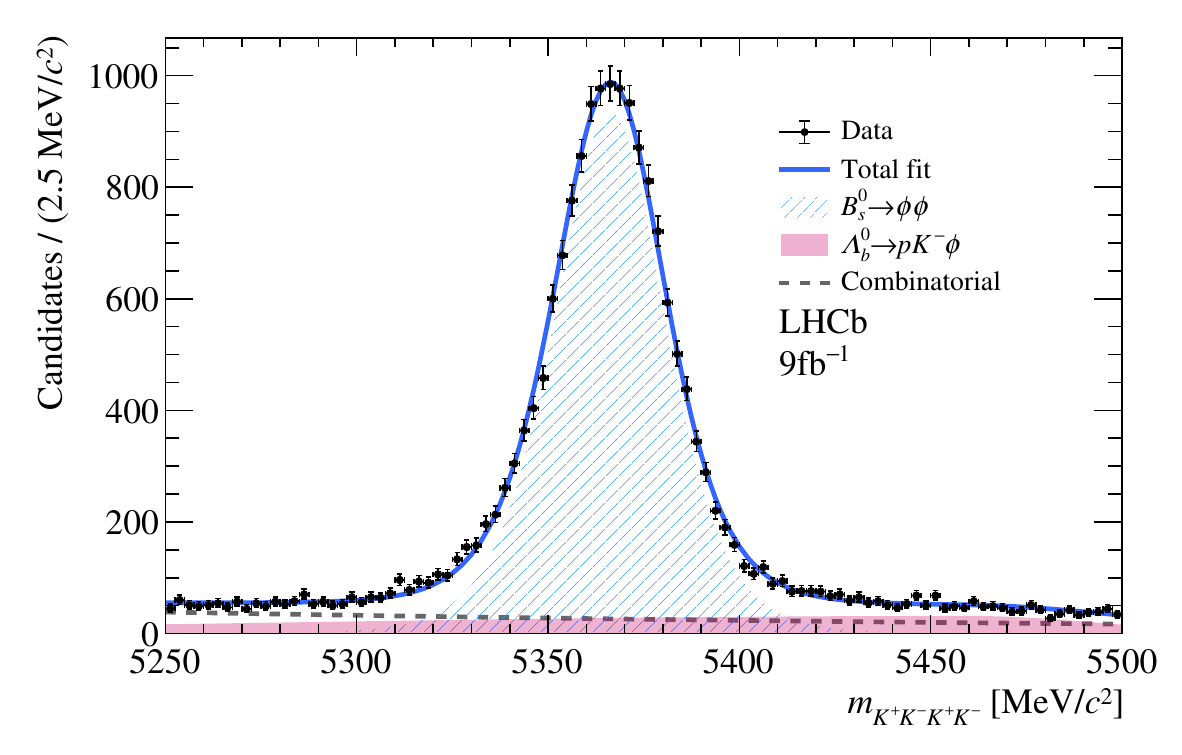}
    \caption{The $\Kp\Km\Kp\Km$ invariant-mass distribution for the normalisation mode for the complete dataset, summing over the four categories. The simultaneous fit result is shown.}
    \label{fig:4kmass_total}
\end{figure}

The $\Kp\Km\Kp\Km$ invariant-mass distribution for the normalisation mode is shown in Fig.~\ref{fig:4kmass_total}. The yields obtained for the four categories are summarised in Table~\ref{tab:4Kyield_and_ratio}. The branching fraction of $\decay{\Bs}{\phi \eta'}$ relative to the $\decay{\Bs}{\phi \phi}$ mode decay is determined for each category using the relation
\begin{equation}
{\cal R}\equiv\frac{\mathcal{B}(\decay{\Bs}{\phi \eta'}) }{\mathcal{B}(\decay{\Bs}{\phi \phi})} = \frac{N_{\phi\eta'} \cdot \mathcal{B}(\decay{\phi}{\Kp \Km})}{N_{\phi\phi} \cdot \mathcal{B}(\decay{\eta'}{\rhoz \gamma})} \cdot \frac{\epsilon_{\phi\phi}}{\epsilon_{\phi\eta'}},
\label{eq:R}
\end{equation}
where $N_{\phi\eta'}$ and $N_{\phi\phi}$ are the fitted yields.

\begin{table}[tb!]
    \centering
     \caption{Fitted yields of normalisation mode, together with the ratio of selection efficiencies in Run~1 and Run~2.}
    \begin{tabular}{l r@{\:$\pm$\:}l r@{\:$\pm$\:}l}
    \hline
              & \multicolumn{2}{c}{$N_{\phi\phi}$} & \multicolumn{2}{c}{$\frac{\epsilon_{\phi\phi}}{\epsilon_{\phi\eta'}}$} \\
    \hline
         Run~1 TIS & $1739$ & $48$ & $11.0$ & $0.4$ \\
         Run~1 TOS & $1285$ & $40$ & $9.7$ & $1.0$ \\
         Run~2 TIS & $6510$ & $94$ & $6.7$ & $0.2$ \\
         Run~2 TOS & $5057$ & $83$ & $6.4$ & $0.2$ \\
    \hline
    \end{tabular}
    \label{tab:4Kyield_and_ratio}
\end{table}

The sources of systematic uncertainty are summarised in Table~\ref{tab:sys}. The main source of experimental uncertainty comes from the efficiency of the hardware trigger, which is determined using data \cite{LHCb-PUB-2014-039}. An additional systematic uncertainty is assigned to account for possible differences in the photon reconstruction efficiency between data and simulation. The simulated PID response is calibrated to better match the data using control samples \cite{LHCb-PUB-2016-021}, with variations in the calibration reference model propagated to $\cal R$.

The impact of the $\decay{\Bs}{\phi \eta'}$ fit model choice on the yield is evaluated using pseudoexperiments, where an alternative model is selected instead. Specifically, the signal shape is described by a skewed Student's t-distribution, and the combinatorial background is described by a polynomial. Pseudoexperiments are generated with both the original and alternative models and fitted using the same procedure as the data. The yields of the signal in pseudoexperiments are compared, with the observed difference taken as the associated uncertainty. The same procedure is used for the normalisation model, using a DSCB function as the signal shape, and the yield is recalculated.

Since both modes contain four tracks, uncertainties originating from the knowledge of the track reconstruction cancel in the efficiency ratio $\frac{\epsilon_{\phi\phi}}{\epsilon_{\phi\eta'}}$. The total amount of material in the detector is known with an accuracy of $10 \%$ \cite{LHCb-DP-2014-002,Fave:2008zz}. Due to the difference in the hadronic-interaction probabilities for kaons and pions, this leads to a $0.7 \%$ uncertainty on $\cal R$. The final state of the signal decay mode is a $\CP$-even eigenstate; therefore, assuming \CP conservation, the lifetime of the $\Bs$ is assumed to be $\tau_L=1.427 \pm 0.007 \, \ps$ \cite{HFLAV23}. Using the alternative value $\tau = 1/\Gamma_s = 1.516 \pm 0.006 \, \ps $ \cite{HFLAV23} changes $\frac{\epsilon_{\phi\phi}}{\epsilon_{\phi\eta'}}$ by $2.8\%$, which is taken as a systematic uncertainty. Finally, a further $1.7  \%$ uncertainty arises from the knowledge of the $\decay{\eta'}{\rhoz \gamma}$ and $\decay{\phi}{\Kp \Km}$ branching fractions.

\begin{figure}[htb!]
    \centering
    \includegraphics[width=0.8\linewidth]{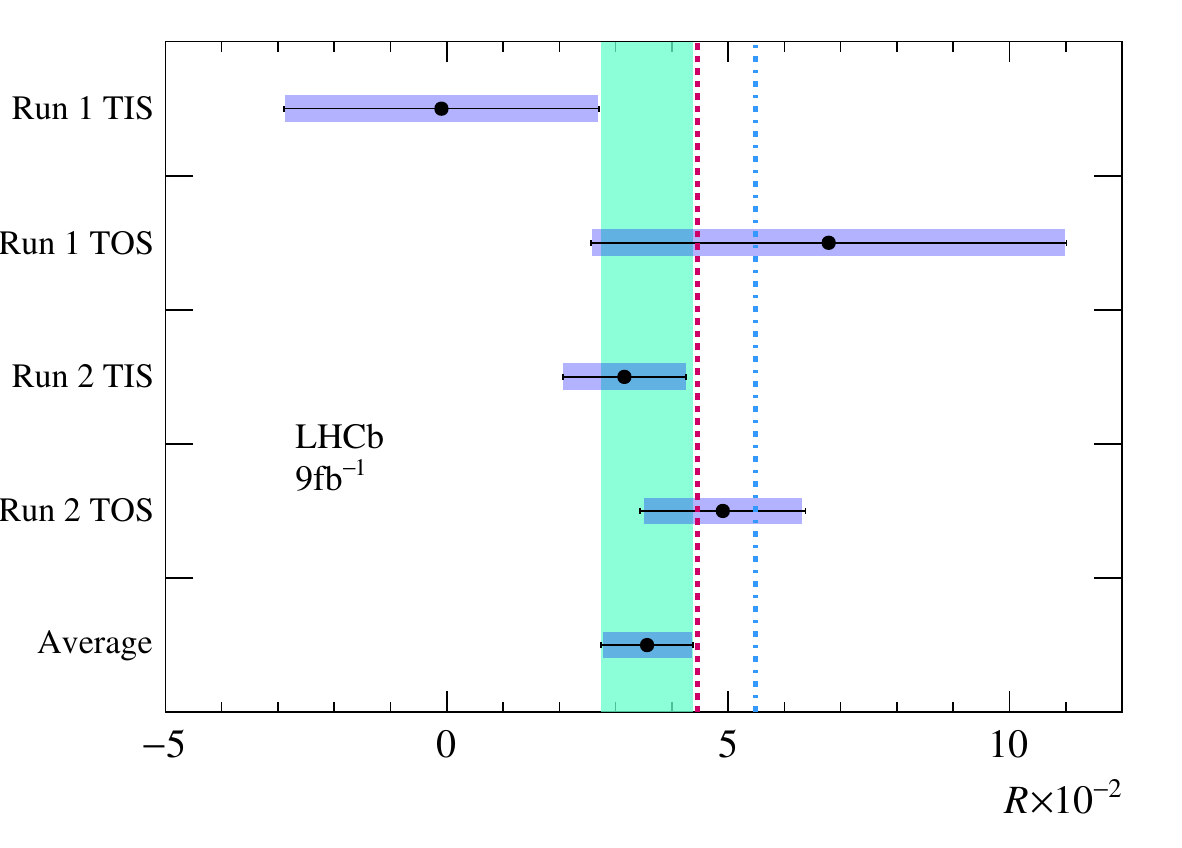}
    \caption{Ratio of branching fraction $\cal R$ for each of the categories and their weighted average, compared to the upper limit at 90\% (red dashed line) and 95\% (cyan dot-dashed line) confidence level previously obtained with the Run~1 dataset \cite{LHCB-PAPER-2016-060}. The shaded area on the error bar represents the statistical, while the full error bar represents the total uncertainty. The green band shows the average value in this analysis.}
    \label{fig:R_summary}
\end{figure}

\begin{table}[tb!]
    \centering
     \caption{Sources of systematic uncertainty on the branching-fraction ratio ${\cal R}$. Relative values for Run 1 and Run 2 are shown in the first two columns. The third column indicates whether the uncertainties are considered as correlated in the averaging procedure.}
    \label{tab:sys}
    \setlength{\tabcolsep}{3mm}{
    \begin{tabular}{lccc}
        \hline
         Sources of uncertainties & Run~1 [\%] & Run~2 [\%] & Correlated \\
         \hline
         Hardware Trigger efficiency  & $9.8$ & $1.6$& No \\
         Photon reconstruction efficiency&  $3.8$ & $2.4$& No \\     
         PID calibration & $4.2$ & $1.3$& No \\
         Offline selection &  $3.2$ & $2.5$& No \\
         Fitting model (signal mode) & $1.8$  & $4.8$& No \\
         Fitting model (control mode) &  $1.8$ & $4.6$& No \\
         Hadronic interaction & $0.7$ & $0.7$& Yes\\
         $B^0_s$ lifetime & $2.8$ & $2.8$& Yes\\
         \hline
         Branching fraction of $\phi$ and $\eta'$ decays & $1.7$& $1.7$& Yes\\
         \hline
    \end{tabular}
    }
\end{table}
For each category, the yield obtained from the simultaneous fit (shown in Table~\ref{tab:phieta_yield}) is converted into a ratio of branching fractions using Eq.~\ref{eq:R}. A weighted average of the four values, which properly takes into account the correlations of uncertainties between them, is then calculated. This procedure gives ${\cal R} =(3.56\pm 0.79\pm 0.18\pm 0.06)\times10^{-2}$,
where the first uncertainty is statistical, the second is systematic, and the last is due to the external branching fractions used. The ratio of branching fractions for the four categories are shown in Fig.~\ref{fig:R_summary}, together with the weighted average value. The average ratio of the branching fractions between the signal and normalisation modes obtained in this analysis is below the limit set by the Run~1 analysis \cite{LHCB-PAPER-2016-060}. 

\section{Summary}
\label{sec:summary}
Using a dataset corresponding to an integrated luminosity of approximately $9 \invfb$ collected in $pp$ collisions by the LHCb experiment between 2011 and 2018, $3.5 \sigma$ evidence for the decay $\decay{\Bs}{\phi \eta'}$ is found. The branching ratio relative to the $\decay{\Bs}{\phi \phi}$ decay is determined to be 
\begin{equation*}
    {\cal R}=(3.56 \pm 0.79\pm 0.18\pm 0.06)\times10^{-2},
\end{equation*}
where the first uncertainty is statistical, the second is systematic, and the third arises due to the external branching fractions used. Using the known value of the branching fraction of the normalisation mode $\mathcal{B}(\decay{\Bs}{\phi \phi}) = (1.84 \pm 0.14 ) \times 10^{-5}$ \cite{PDG2024}, the branching fraction 
\begin{equation*}
    \mathcal{B}(\decay{\Bs}{\phi\eta'})=(0.66 \pm 0.15 \pm 0.03 \pm 0.02) \times 10^{-6},
\end{equation*}
is obtained. This value is within the range $(0.05\--20)\times10^{-6}$ of theoretical predictions in Ref. \cite{Beneke:2003zv,Cheng:2009mu,Ali:2007ff,Chen:2007qm, Wang:2008rk, PhysRevD.91.014011, Zhou:2016jkv}, and helps to constrain QCD models for charmless $b$-hadron decays to a V-P/P-V state. It supersedes the result presented in Ref.~\cite{LHCB-PAPER-2016-060}. Further improvements to the knowledge of this mode and similar decays will be made using the larger dataset collected during Run 3 by the upgraded LHCb detector with an improved trigger for hadronic modes.

\section*{Acknowledgements}
%
%
\noindent We express our gratitude to our colleagues in the CERN
accelerator departments for the excellent performance of the LHC. We
thank the technical and administrative staff at the LHCb
institutes.
We acknowledge support from CERN and from the national agencies:
ARC (Australia);
CAPES, CNPq, FAPERJ and FINEP (Brazil); 
MOST and NSFC (China); 
CNRS/IN2P3 and CEA (France);  
BMFTR, DFG and MPG (Germany);
INFN (Italy); 
NWO (Netherlands); 
MNiSW and NCN (Poland); 
MEC/IFA (Romania); 
MICIU and AEI (Spain);
SNSF and SER (Switzerland); 
NASU (Ukraine); 
STFC (United Kingdom); 
DOE NP and NSF (USA).
We acknowledge the computing resources that are provided by ARDC (Australia), 
CBPF (Brazil),
CERN, 
IHEP and LZU (China),
IN2P3 (France), 
KIT and DESY (Germany), 
INFN (Italy), 
SURF (Netherlands),
Polish WLCG (Poland),
IFIN-HH (Romania), 
PIC (Spain), CSCS (Switzerland), 
GridPP (United Kingdom),
and NSF (USA).  
We are indebted to the communities behind the multiple open-source
software packages on which we depend.
Individual groups or members have received support from
RTP (Australia), 
FWO Odysseus grant G0ASD25N (Belgium), 
Key Research Program of Frontier Sciences of CAS, CAS PIFI, CAS CCEPP (China); 
Minciencias (Colombia);
EPLANET, Marie Sk\l{}odowska-Curie Actions, ERC and NextGenerationEU (European Union);
A*MIDEX, ANR, IPhU and Labex P2IO, and R\'{e}gion Auvergne-Rh\^{o}ne-Alpes (France);
Alexander-von-Humboldt Foundation (Germany);
ICSC (Italy); 
Severo Ochoa and Mar\'ia de Maeztu Units of Excellence, GVA, XuntaGal, GENCAT, InTalent-Inditex and Prog.~Atracci\'on Talento CM (Spain);
the Leverhulme Trust, the Royal Society and UKRI (United Kingdom).


\clearpage
\addcontentsline{toc}{section}{References}
\bibliographystyle{LHCb}
\bibliography{main,standard,LHCb-PAPER,LHCb-CONF,LHCb-DP,LHCb-TDR}

@article{LHCb-DP-2023-003,
      author         = "Roel Aaij and others",
      collaboration  = "LHCb collaboration",
      title         = "{Momentum scale calibration of the LHCb spectrometer}",
      eprint         = "2312.01772",
      archivePrefix  = "arXiv",
      primaryClass   = "hep-ex",
      report         = "LHCb-DP-2023-003, CERN-EP-2023-275",
      year           = "2024",
      journal        = "JINST",
      volume         = "19",
      pages          = "P02008",
      doi            = "10.1088/1748-0221/19/02/P02008",
      }

@article{LHCb-DP-2019-001,
      author         = "Roel Aaij and others",
      title         = "{Design and performance of the LHCb trigger and full real-time  reconstruction in Run 2 of the LHC}",
      eprint         = "1812.10790",
      archivePrefix  = "arXiv",
      primaryClass   = "hep-ex",
      number         = "LHCb-DP-2019-001",
      year           = "2019",
      journal        = "JINST",
      volume         = "14",
      pages          = "P04013",
      doi            = "10.1088/1748-0221/14/04/P04013",
      }

@article{LHCb-DP-2014-002,
      author         = "Aaij, R. and others",
      title          = "{LHCb detector performance}",
      collaboration  = "LHCb collaboration",
      journal        = "Int. J. Mod. Phys.",
      volume         = "A30",
      pages          = "1530022",
      doi            = "10.1142/S0217751X15300227",
      year           = "2015",
      eprint         = "1412.6352",
      archivePrefix  = "arXiv",
      primaryClass   = "hep-ex",
      report         = "LHCB-DP-2014-002, CERN-PH-EP-2014-290",
}

@article{LHCb-DP-2014-001,
      author         = "Aaij, R. and others",
      title          = "{Performance of the LHCb Vertex Locator}",
      journal        = "JINST",
      volume         = "9",
      pages          = "P09007",
      doi            = "10.1088/1748-0221/9/09/P09007",
      year           = "2014",
      eprint         = "1405.7808",
      archivePrefix  = "arXiv",
      primaryClass   = "physics.ins-det",
      report         = "LHCB-DP-2014-001",
}

@article{LHCb-DP-2012-004,
      author         = "Aaij, R. and others",
      title          = "{The \lhcb trigger and its performance in 2011}",
      journal        = "JINST",
      volume         = "8",
      pages          = "P04022",
      doi            = "10.1088/1748-0221/8/04/P04022",
      year           = "2013",
      eprint         = "1211.3055",
      archivePrefix  = "arXiv",
      primaryClass   = "hep-ex",
      report         = "LHCb-DP-2012-004",
}

@article{LHCb-DP-2008-001,
      author         = "Alves~Jr., A. A. and others",
      title          = "{The \lhcb detector at the LHC}",
      collaboration  = "LHCb collaboration",
      journal        = "JINST",
      volume         = "3",
      pages          = "S08005",
      doi            = "10.1088/1748-0221/3/08/S08005",
      year           = "2008",
      number         = "LHCb-DP-2008-001",
}

@article{LHCb-PAPER-2025-018,
      author         = "Aaij, R. and others",
      title          = "{Search for the decay $B^0 \to \phi \phi$}",
      collaboration  = "LHCb collaboration",
      report         = "{LHCb-PAPER-2025-018, CERN-EP-2025-150}",
      eprint         = "2507.20945",
      archivePrefix  = "arXiv",
      primaryClass   = "hep-ex",
      journal        = "{JHEP}",
      volume         = "12",
      year           = "2025",
      pages          = "26",
      doi            = "10.1007/JHEP12(2025)026"
}

@article{LHCb-PAPER-2016-060,
      author         = "Aaij, R. and others",
      title          = "{Search for the \mbox{\decay{\Bs}{\etapr\phiz}} decay}",
      collaboration  = "LHCb collaboration",
      year           = "2017",
      journal        = "JHEP",
      volume         = "05",
      pages          = "158",
      doi            = "10.1007/JHEP05(2017)158",
      report         = "{LHCb-PAPER-2016-060 CERN-EP-2016-304}",
      eprint         = "1612.08110",
      archivePrefix  = "arXiv",
      primaryClass   = "hep-ex",
}

@article{LHCb-PAPER-2016-054,
      author         = "Aaij, R. and others",
      title          = "{Measurement of the \Bpm production asymmetry 
                         and the \CP asymmetry in 
                         \mbox{\decay{\Bpm}{\jpsi \Kpm}} decays}",
      collaboration  = "LHCb collaboration",
      year           = "2017",
      journal        = "Phys. Rev.",
      volume         = "D95",
      pages          = "052005",
      doi            = "10.1103/PhysRevD.95.052005",
      report         = "{LHCb-PAPER-2016-054 CERN-EP-2016-325}",
      eprint         = "1701.05501",
      archivePrefix  = "arXiv",
      primaryClass   = "hep-ex",
}

@lhcbreport{LHCb-TDR-001,
      collaboration = "LHCb collaboration",
      title         = "{LHCb magnet: Technical Design Report}",
      institution   = "CERN",
      address       = "Geneva",
      number        = "{CERN-LHCC-2000-007}",
      year          = "2000",
}

@article{mnsc.44.12.1650,
author = {Theodossiou, Panayiotis},
title = "{Financial data and the skewed generalized T distribution}",
journal = {Manag. Sci.},
volume = {44},
number = {12-part-1},
pages = {1650-1661},
year = {1998},
doi = {10.1287/mnsc.44.12.1650},
URL = { 
        https://pubsonline.informs.org/doi/abs/10.1287/mnsc.44.12.1650}
}

@article{Beneke:2003zv,
    author = "Beneke, Martin and Neubert, Matthias",
    title = "{QCD factorization for $B\to PP$ and $B\to PV$ decays}",
    eprint = "hep-ph/0308039",
    archivePrefix = "arXiv",
    reportNumber = "CLNS-03-1835, PITHA-03-06",
    doi = "10.1016/j.nuclphysb.2003.09.026",
    journal = "Nucl. Phys.",
    volume = "B675",
    pages = "333--415",
    year = "2003"
}

@article{Ali:2007ff,
    author = "Ali, Ahmed and Kramer, Gustav and Li, Ying and Lu, Cai-Dian and Shen, Yue-Long and Wang, Wei and Wang, Yu-Ming",
    title = "{Charmless nonleptonic $B_s$ decays to $PP$, $PV$ and $VV$ final states in the pQCD approach}",
    eprint = "hep-ph/0703162",
    archivePrefix = "arXiv",
    reportNumber = "DESY-07-021",
    doi = "10.1103/PhysRevD.76.074018",
    journal = "Phys. Rev.",
    volume = "D76",
    pages = "074018",
    year = "2007"
}

@article{Chen:2007qm,
    author = "Chen, Xin-fen and Guo, Dong-qin and Xiao, Zhen-jun",
    title = "{$B_s\to(\rho,\omega,\phi)\eta'$ decays in the perturbative QCD approach}",
    eprint = "hep-ph/0701146",
    archivePrefix = "arXiv",
    reportNumber = "NJNU-TH-07-02",
    month = "1",
    year = "2007"
}

@article{Wang:2008rk,
    author = "Wang, Wei and others",
    title = "{Charmless two-body $B_s\to VP$ decays in soft collinear effective theory}",
    eprint = "0801.3123",
    archivePrefix = "arXiv",
    primaryClass = "hep-ph",
    doi = "10.1103/PhysRevD.78.034011",
    journal = "Phys. Rev.",
    volume = "D78",
    pages = "034011",
    year = "2008"
}

@lhcbreport{Fave:2008zz,
     author        = "Fave, V",
     collaboration = "LHCb",
     institution   = "CERN",
     title         = "{Estimation of the material budget of the Inner Tracker}",
     address       = "Geneva",
     year          = "2008",
     number        = "LHCb-2008-054, CERN-LHCb-2008-054",
     url           = "https://cds.cern.ch/record/1134016",
}

@article{Cheng:2009mu,
    author = "Cheng, Hai-Yang and Chua, Chun-Khiang",
    title = "{{QCD} factorization for charmless hadronic $B_s$ decays revisited}",
    eprint = "0910.5237",
    archivePrefix = "arXiv",
    primaryClass = "hep-ph",
    doi = "10.1103/PhysRevD.80.114026",
    journal = "Phys. Rev.",
    volume = "D80",
    pages = "114026",
    year = "2009"
}

@article{jbes.2009.06161,
author = {Christian Hansen and James B. McDonald and Whitney K. Newey},
title = "{Instrumental variables estimation with flexible distributions}",
journal = {J. Bus. Econ. Stat.},
volume = {28},
number = {1},
pages = {13-25},
year  = {2010},
publisher = {Taylor & Francis},
doi = {10.1198/jbes.2009.06161},
URL = {https://doi.org/10.1198/jbes.2009.06161}
}

@article{PhysRevD.91.014011,
  title = "{Updating $B\ensuremath{\rightarrow}PP, VP$ decays in the framework of flavor symmetry}",
  author = {Cheng, Hai-Yang and Chiang, Cheng-Wei and Kuo, An-Li},
  journal = {Phys. Rev.},
  volume = {D91},
  issue = {1},
  pages = {014011},
  numpages = {22},
  year = {2015},
  month = {Jan},
  publisher = {American Physical Society},
  doi = {10.1103/PhysRevD.91.014011},
  url = {https://link.aps.org/doi/10.1103/PhysRevD.91.014011},
  eprint={1409.5026},
  archivePrefix={arXiv},
  primaryClass={hep-ph},
}

@article{Zhou:2016jkv,
    author = {Zhou, Si-Hong and Zhang, Qi-An and Lyu, Wei-Ran and L{\"u}, Cai-Dian},
    title = "{Analysis of charmless two-body B decays in factorization assisted topological amplitude approach}",
    eprint = "1608.02819",
    archivePrefix = "arXiv",
    primaryClass = "hep-ph",
    doi = "10.1140/epjc/s10052-017-4685-0",
    journal = "Eur. Phys. J.",
    volume = "C77",
    number = "2",
    pages = "125",
    year = "2017"
}

@inproceedings{Chen:2016:XST:2939672.2939785,
 author = {Chen, Tianqi and Guestrin, Carlos},
 title = {{XGBoost}: A Scalable Tree Boosting System},
 booktitle = {Proceedings of the 22nd ACM SIGKDD International Conference on Knowledge Discovery and Data Mining},
 series = {KDD '16},
 year = {2016},
 isbn = {978-1-4503-4232-2},
 location = {San Francisco, California, USA},
 pages = {785--794},
 numpages = {10},
 url = {http://doi.acm.org/10.1145/2939672.2939785},
 doi = {10.1145/2939672.2939785},
 acmid = {2939785},
 publisher = {ACM},
 address = {New York, NY, USA},
}

@phdthesis{Timothy2018,
  author       = {Timothy Williams},
  title        = {First observation of the charmless beauty decay $\decay{\Lb}{p \Km\eta'}$},
  school       = {University of Birmingham},
  year         = {2018},
  address      = {Birmingham, United Kingdom},
  month        = {May},
}

@article{PDG2024,
     author    = "Navas, S. and others",
    collaboration = "Particle Data Group",
     title     = "{\href{http://pdg.lbl.gov/}{Review of particle physics}}",
     journal   = "Phys. Rev.",
     year = {2024},
     number = {8},
     volume      = "D110",
     pages     = "030001",
     doi = "10.1103/PhysRevD.110.030001"
}

@article{LHCb-PROC-2015-018,
      author        = "Likhomanenko, T. and others",
      title         = "{LHCb topological trigger reoptimization}",
      journal       = "J. Phys. Conf. Ser.",
      volume        = "664",
      pages         = "082025",
      doi           = "10.1088/1742-6596/664/8/082025",
      month         = "Oct",
      year          = "2015",
      reportNumber  = "LHCb-PROC-2015-018",
      eprint        = "1510.00572",
      archivePrefix = "arXiv"
}

@article{LHCb-PROC-2011-006,
  author="Clemencic, M and others",
  title="{The \lhcb simulation application, Gauss: Design, evolution and experience}",
  journal="J. Phys. Conf. Ser.",
  volume={331},
  pages={032023},
  doi={10.1088/1742-6596/331/3/032023},
  year={2011},
}

@article{LHCb-PROC-2010-056,
      author         = "Belyaev, I. and others",
      title          = "{Handling of the generation of primary events
                         in Gauss, the LHCb simulation framework}",
      journal="J. Phys. Conf. Ser.",
      volume={331},
      pages={032047},
      doi={10.1088/1742-6596/331/3/032047},
      year={2011},
}

@article{HFLAV23,
      author         = "Banerjee, S. and others",
      collaboration  = "Heavy Flavor Averaging Group",
      title          = "{Averages of $b$-hadron, $c$-hadron, and $\tau$-lepton properties as of 2023}",
      eprint         = "2411.18639",
      archivePrefix  = "arXiv",
      primaryClass   = "hep-ex",
      year           = "2026",
      journal        = "Phys. Rev.",
      volume         = "D113",
      pages          = "012008",
      doi            = "10.1103/x87q-tld5"
}

@article{Sjostrand:2006za,
      author         = {Sj\"{o}strand, Torbj\"{o}rn and Mrenna, Stephen and Skands, Peter"},
      title          = "{PYTHIA 6.4 physics and manual}",
      journal        = "JHEP",
      volume         = "05",
      pages          = "026",
      doi            = "10.1088/1126-6708/2006/05/026",
      year           = "2006",
      eprint         = "hep-ph/0603175",
      archivePrefix  = "arXiv",
      primaryClass   = "hep-ph",
}

@article{Sjostrand:2007gs,
      author         = {Sj\"{o}strand, Torbj\"{o}rn and Mrenna, Stephen and
                        Skands, Peter"},
      title          = "{A brief introduction to PYTHIA 8.1}",
      journal        = "Comput. Phys. Commun.",
      volume         = "178",
      pages          = "852-867",
      doi            = "10.1016/j.cpc.2008.01.036",
      year           = "2008",
      eprint         = "0710.3820",
      archivePrefix  = "arXiv",
      primaryClass   = "hep-ph",
      reportNumber   = "CERN-LCGAPP-2007-04, LU-TP-07-28,
                        FERMILAB-PUB-07-512-CD-T",
}

@Article{Agostinelli:2002hh,
     author    = "Agostinelli, S. and others",
 collaboration = "Geant4 collaboration",
     title     = "{Geant4: A simulation toolkit}",
     journal   = "Nucl. Instrum. Meth.",
     volume    = "A506",
     year      = "2003",
     pages     = "250",
     doi       = "10.1016/S0168-9002(03)01368-8",
}

@article{Allison:2006ve,
      author         = "Allison, John and Amako, K. and Apostolakis, J. and
                        Araujo, H. and Dubois, P.A. and others",
 collaboration = "Geant4 collaboration",
      title          = "{Geant4 developments and applications}",
      journal        = "IEEE Trans.Nucl.Sci.",
      volume         = "53",
      pages          = "270",
      doi            = "10.1109/TNS.2006.869826",
      year           = "2006",
      reportNumber   = "SLAC-PUB-11870",
}

@Article{Lange:2001uf,
     author    = "Lange, D. J.",
     title     = "{The EvtGen particle decay simulation package}",
     journal   = "Nucl. Instrum. Meth.",
     volume    = "A462",
     year      = "2001",
     pages     = "152-155",
     doi       = "10.1016/S0168-9002(01)00089-4",
}

@article{davidson2015photos,
	author="Davidson, N. and Przedzinski, T. and Was, Z.",
	title = "{PHOTOS interface in C++: Technical and physics documentation}",
      	eprint={1011.0937},
      	archivePrefix={arXiv},
      	primaryClass={hep-ph},
	journal = {Comput. Phys. Commun.},
	volume = {199},
	pages = {86},
	year = {2016},
	doi = {https://doi.org/10.1016/j.cpc.2015.09.013},
}

@PhdThesis{Skwarnicki:1986xj,
  author    = "Skwarnicki, Tomasz",
  title     = "{A study of the radiative cascade transitions between the
                  Upsilon-prime and Upsilon resonances}",
  school = 	 {Institute of Nuclear Physics, Krakow},
  year = 	 {1986},
  note = 	 "{\href{http://inspirehep.net/record/230779/}{DESY-F31-86-02}}",
}

@article{Cowan:2016tnm,
      author         = "Cowan, G. A. and Craik, D. C. and Needham, M. D.",
      title          = "{RapidSim: an application for the fast simulation of
                        heavy-quark hadron decays}",
      journal        = "Comput. Phys. Commun.",
      volume         = "214",
      year           = "2017",
      pages          = "239-246",
      doi            = "10.1016/j.cpc.2017.01.029",
      eprint         = "1612.07489",
      archivePrefix  = "arXiv",
      primaryClass   = "hep-ex",
}

@article{BBDT,
      author         = "Gligorov, V. V. and Williams, M.",
      title          = "{Efficient, reliable and fast high-level triggering using a bonsai boosted decision tree}",
      journal        = "JINST",
      volume         = "8",
      pages          = "P02013",
      doi            = "10.1088/1748-0221/8/02/P02013",
      year           = "2013",
      eprint         = "1210.6861",
      archivePrefix  = "arXiv",
      primaryClass   = "physics.ins-det",

}

@article{Hulsbergen:2005pu,
      author         = "Hulsbergen, Wouter D.",
      title          = "{Decay chain fitting with a Kalman filter}",
      journal        = "Nucl. Instrum. Meth.",
      volume         = "A552",
      pages          = "566-575",
      doi            = "10.1016/j.nima.2005.06.078",
      year           = "2005",
      eprint         = "physics/0503191",
      archivePrefix  = "arXiv",
      primaryClass   = "physics",
}

@article{Wilks:1938dza,
      author         = "Wilks, S. S.",
      title          = "{The large-sample distribution of the likelihood ratio
                        for testing composite hypotheses}",
      journal        = "Ann. Math.Stat.",
      volume         = "9",
      pages          = "60-62",
      doi            = "10.1214/aoms/1177732360",
      year           = "1938",
}

@article{Punzi:2003bu,
      author         = "Punzi, Giovanni",
      title          = "{Sensitivity of searches for new signals and its
                        optimization}",
      booktitle      = "{Statistical problems in particle physics, astrophysics
                        and cosmology. Proceedings, Conference, PHYSTAT 2003,
                        Stanford, USA, September 8-11, 2003}",
      journal        = "eConf",
      volume         = "C030908",
      year           = "2003",
      pages          = "MODT002",
      eprint         = "physics/0308063",
      archivePrefix  = "arXiv",
      primaryClass   = "physics",
      reportNumber   = "PHYSTAT-2003-MODT002",
}

@article{Kobayashi:1973fv,
      author         = "Kobayashi, Makoto and Maskawa, Toshihide",
      title          = "{\CP-violation in the renormalizable theory of weak
                        interaction}",
      journal        = "Prog. Theor. Phys.",
      volume         = "49",
      year           = "1973",
      pages          = "652-657",
      doi            = "10.1143/PTP.49.652",
      reportNumber   = "KUNS-242",
}

@Misc{mciteplus,
  author = 	 {Shell, Michael},
  title = 	 {Mciteplus: Enhanced multicitations},
  howpublished = {\href{http://www.michaelshell.org/tex/mciteplus/} {http://www.michaelshell.org/tex/mciteplus/}},
}

@lhcbreport{LHCb-PUB-2016-021,
      author        = "Anderlini, Lucio and Contu, Andrea and Jones, Christopher
                       Rob and Malde, Sneha Sirirshkumar and Muller, Dominik and
                       Ogilvy, Stephen and Otalora Goicochea, Juan Martin and
                       Pearce, Alex and Polyakov, Ivan and Qian, Wenbin and
                       Sciascia, Barbara and Vazquez Gomez, Ricardo and Zhang,
                       Yanxi",
      title         = "{The PIDCalib package}",
      number        = "LHCb-PUB-2016-021",
      year          = "2016",
      url           = "http://cds.cern.ch/record/2202412",
}

@lhcbreport{LHCb-PUB-2014-039,
      author        = "Tolk, S and Albrecht, J and Dettori, F and Pellegrino, A",
      title         = "{Data driven trigger efficiency determination at LHCb}",
      number        = "LHCb-PUB-2014-039",
      year          = "2014",
}

@article{Stripping,
      author        = "Grieser, Nathan and Rodrigues, Eduardo and Sahoo, Niladri and Sheng, 
                       Shuqi and Skidmore, Nicole and Smith, Mark",
      title         = "{The LHCb stripping project: Sustainable legacy data processing for high-energy physics}",
      eprint        = "2509.05294",
      archivePrefix = "arXiv",
      primaryClass  = "hep-ex",
      year          = "2025",
      journal       = "Comput. Softw. Big. Sci.",
      volume        = "9",
      pages         = "21",
      doi           = "10.1007/s41781-025-00151-6"
}

\clearpage
\centerline
{\large\bf LHCb collaboration}
\begin
{flushleft}
\small
R.~Aaij$^{38}$\lhcborcid{0000-0003-0533-1952},
M.~Abdelfatah$^{69}$,
A.S.W.~Abdelmotteleb$^{57}$\lhcborcid{0000-0001-7905-0542},
C.~Abellan~Beteta$^{51}$\lhcborcid{0009-0009-0869-6798},
F.~Abudin\'en$^{59}$\lhcborcid{0000-0002-6737-3528},
T.~Ackernley$^{61}$\lhcborcid{0000-0002-5951-3498},
A.A.~Adefisoye$^{69}$\lhcborcid{0000-0003-2448-1550},
B.~Adeva$^{47}$\lhcborcid{0000-0001-9756-3712},
M.~Adinolfi$^{55}$\lhcborcid{0000-0002-1326-1264},
P.~Adlarson$^{87}$\lhcborcid{0000-0001-6280-3851}
% ...

{\footnotesize \it

% ...
$^{38}$Nikhef National Institute for Subatomic Physics, Amsterdam, Netherlands\\
% ...
$^{47}$Instituto Galego de F{\'\i}sica de Altas Enerx{\'\i}as (IGFAE), Universidade de Santiago de Compostela, Santiago de Compostela, Spain\\
% ...
$^{51}$Physik-Institut, Universit{\"a}t Z{\"u}rich, Z{\"u}rich, Switzerland\\
% ...
$^{55}$H.H. Wills Physics Laboratory, University of Bristol, Bristol, United Kingdom\\
% ...
$^{57}$Department of Physics, University of Warwick, Coventry, United Kingdom\\
% ...
$^{59}$School of Physics and Astronomy, University of Edinburgh, Edinburgh, United Kingdom\\
$^{60}$School of Physics and Astronomy, University of Glasgow, Glasgow, United Kingdom\\
$^{61}$Oliver Lodge Laboratory, University of Liverpool, Liverpool, United Kingdom\\
% ...
$^{69}$Syracuse University, Syracuse, NY, United States\\
% ...
$^{87}$Department of Physics and Astronomy, Uppsala University, Uppsala, Sweden, associated to $^{60}$\\
% ...
}
\end{flushleft}

\end{document}